\documentclass[a4paper,11pt,aps,prd,nofootinbib,superscriptaddress]{revtex4-1}
\usepackage[margin=0.9in]{geometry}

\usepackage{slashed}
\usepackage{caption}
\usepackage{amsmath,amssymb,graphicx,xcolor,mathtools}
\usepackage[colorlinks=true,linktocpage=true,linkcolor=red,citecolor=blue]{hyperref}
\numberwithin{equation}{section}
\allowdisplaybreaks

\def\be {\begin{equation}}
\def\ee {\end{equation}}
\def\bea {\begin{eqnarray}}
\def\eea {\end{eqnarray}}
\def\bc {\begin{center}}
\def\ec {\end{center}}
\def\nn {\nonumber}
\def\gm {\gamma}
\def\de {\delta}
\def\eps {\epsilon}

\def\wt {\widetilde}
\def\la {\langle}
\def\ra {\rangle}
\def\lrw {\leftarrow}
\def\mn {\mu\nu}
\def\th {\theta}
\def\ov {\overline}
\def\sg {\sigma}
\def\om {\omega}

\def\pp {\perp}
\def\pl {\shortparallel}
\DeclareMathOperator{\sech}{sech}

\begin{document}

\title{Effect of magnetic field on dilepton production in a hot plasma}
\author{Aritra Bandyopadhyay}
\author{S. Mallik}
\affiliation{
Theory Division, Saha Institute of Nuclear Physics, HBNI \\
1/AF, Bidhannagar, Kolkata 700064, India.
}
\email{aritra.bandyopadhyay@saha.ac.in}
\email{mallik@theory.saha.ernet.in}
\date{\today}

\begin{abstract}
Noncentral collision of heavy ions can generate large magnetic field in
its neighbourhood. We describe a method to calculate the effect of this
field on the dilepton emission rate from the colliding region, when it
reaches thermal equilibrium. It is calculated in the real time method of 
thermal field theory. We find that the rate is affected significantly only 
for lower momenta of dileptons.
\end{abstract}

\maketitle

\section{Introduction}

An important probe into the dynamics of heavy ion collisions is the detection 
of dilepton production in the process. Accordingly this topic has been
investigated in detail~\cite{McLerran1,Weldon,Rapp1,Tserruya,Rapp2,Vujanovic}. 
Here we study the change in the production rate due to the magnetic field, 
which is produced in non-central collision of individual events~
\cite{Kharzeev,Skokov,Bzdak,McLerran4,Tuchin1,Tuchin2,Tuchin3,Sadooghi,Bandyo}.

According to the present understanding, the two nuclei colliding at ultra 
relativistic energies appear as two sheets of Color Glass Condensate~
\cite{McLerran2}. Very shortly after collision a strongly interacting Quark 
Gluon system, called Glasma, is formed, which is out of thermal equilibrium~
\cite{Baier,Blaizot1,Blaizot2,McLerran3}. After thermalization it gives rise 
to the Quark Gluon plasma (QGP) phase. Finally it evolves into a hadron gas. 
Dileptons are produced in all these phases. In this work we address this 
production in the QGP phase. 

The effect of magnetic field in different processes arises through the altered 
propagation of particles in this field. A non-perturbative, gauge covariant 
expression for the Dirac propagator in an external electromagnetic field was 
derived long ago by Schwinger in an elegant way, using a proper-time parameter~
\cite{Schwinger}. It has since been rederived and applied to many 
processes~\cite{Ritus,Tsai1,Tsai2,rojas,Gusynin,Hattori,Ayala1,Shovkovy}. In 
the present work we expand the exact propagator in powers of the magnetic field. 

The dilepton production rate is given in terms of the (imaginary part of the) 
thermal two-point current correlation function. The latter involves the thermal 
propagator for quarks in the magnetic field. In contrast to the oft-used
imaginary time formulation of thermal field theory~
\cite{Matsubara,Pisarski,Kapusta,Bellac}, we shall use the real time formulation
~\cite{Niemi,Kobes,Mallik}. The advantage is that we do not have frequency sums 
for the propagators, but at the cost of dealing with $2\times 2$ matrices for them  
in the intermediate stage of calculation. The matrices admit spectral
representations, just like the vacuum propagators, which we shall use in
calculating the thermal correlation function.  

In Section II we write the dilepton rate formula and describe our method to
evaluate it. In Section III we outline Schwinger's construction of spinor
propagator in magnetic field, leading to its spectral representation. In
Section IV we then calculate the thermal two-point correlation function of
currents, present in the rate formula. Finally Section \ref{discussion} 
contains the numerical results and discussion. 

\section{Formulation}
\label{form}

The transition amplitude (Fig 1) from an initial state $I$, composed of quarks and 
gluons to a final state $F$ of similar composition, along with the emission of a 
dilepton $l(p,\sg)$ and $\bar{l}(p',\sg')$ of momenta $p$ and $p'$ 
and $z$-component of spin $\sg$ and $\sg'$ is 
\bea
\la F, l(p,\sg), \bar{l}(p',\sg')| S | I\ra.
\eea
Here the scattering matrix operator $S$ is given by the interaction Lagrangian
\bea
\mathcal{L}_{\mathrm{int}} = -e \left(j^\mu(x)+J^\mu(x)\right)A_\mu(x),
\eea
of lepton and quark currents 
\bea
j^\mu(x) = \bar{\psi}(x)\gm^\mu\psi(x),~~J^\mu(x) = \frac{2}{3} \bar{u}(x)\gm^\mu u(x) 
- \frac{1}{3} \bar{d}(x)\gm^\mu d(x), 
\eea
coupled to the electromagnetic field $A_\mu(x)$. We assume the initial state to be 
thermal and look for inclusive probability. Then if $N$ is the dilepton emission rate 
per unit volume, we get, after some calculation \cite{Mallik} 
\bea
\frac{d^4N}{d^4q} = \frac{\alpha^2}{6\pi^3q^2}e^{-\beta q_0}
\left(-g^{\mn}M^+_{\mn}\right),
\label{dileptonrate}
\eea
where $q=p+p'$ is the dilepton momentum and $M^+_{\mn}(q)$ is a thermal 
two-point function of the quark current 
\bea
M^+_{\mn}(q) = \int d^4x e^{iq\cdot x}\la J_\mu(x)J_\nu(0)\ra.
\eea
Here the symbol $\la O\ra$ stands for ensemble average of the operator $O$ at
temperature $1/\beta$,
\bea
\la O\ra = \mathrm{Tr} (e^{-\beta H}O) / \mathrm{Tr} e^{-\beta H}. 
\eea

\begin{center}
\begin{figure}[tbh]
\begin{center}
\includegraphics[scale=0.5]{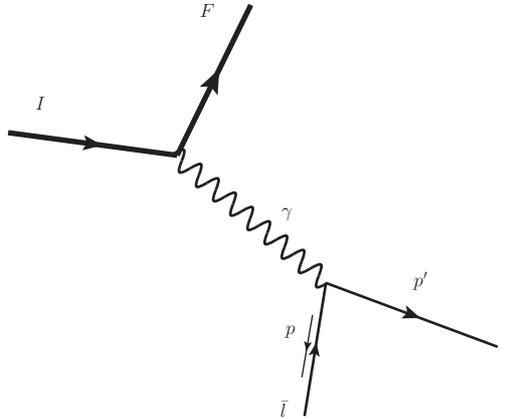} 
\caption{Dilepton production amplitude in QGP phase. The states I and F consist of 
quarks and gluons, while $l\bar{l}$ is a dilepton. The weavy line corresponds to photon.}
\label{dil_pro}
\end{center}
\end{figure}
\end{center}

\begin{center}
\begin{figure}[tbh]
\begin{center}
\includegraphics[scale=0.5]{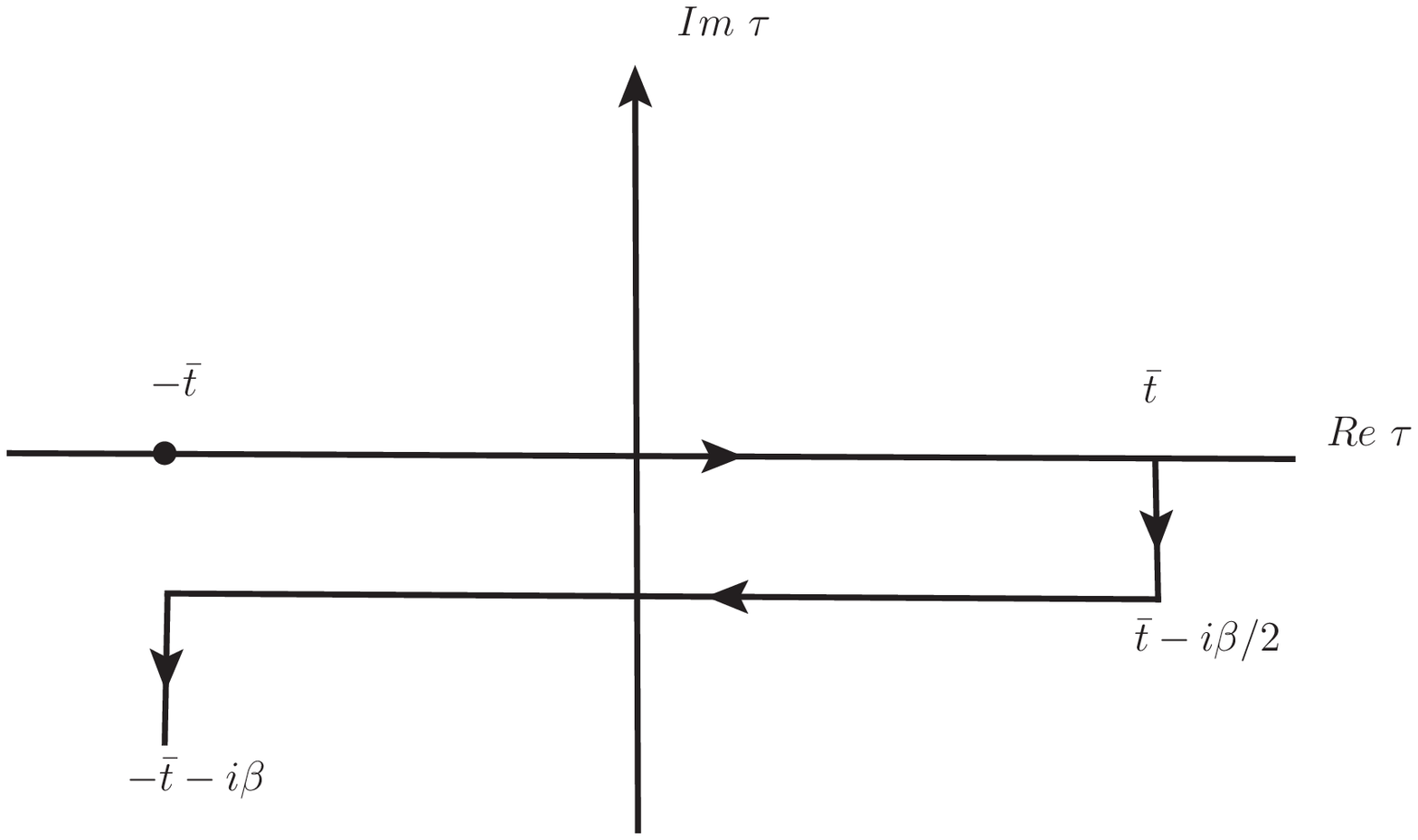} 
\caption{The time contour in complex $\tau$ plane with $\overline{t}\to\infty$.}
\label{rtf_con}
\end{center}
\end{figure}
\end{center}

We briefly review how $M^+_{\mn}(q)$ may be obtained in the real time thermal 
field theory \cite{Mallik}. We start with the time contour of Fig. \ref{rtf_con} and 
define the time-ordered two-point function $M_{\mn}(x,x')$ as
\bea
M_{\mn}(x,x') = \Theta_c(\tau-\tau') i \la J_\mu(x)J_\nu(x')\ra + 
\Theta_c(\tau'-\tau) i \la J_\nu(x')J_\mu(x)\ra.
\label{contourform}
\eea
where $x=(\tau,\vec{x}),~x'=(\tau',\vec{x}')$ with  the 'times' $\tau$ and $\tau'$ 
on the contour shown in Fig 2. The subscript $c$ on the $\Theta$-functions
refers to contour ordering. Beginning with the spatial Fourier transform, one can 
show that the vertical segments of the time contour does not contribute. Then the 
two-point function may be put in the form of a $2\times 2$ matrix, which can be 
diagonalized with essentially one diagonal element
\bea
\ov{M}_{\mn}(q) = \int\limits_{-\infty}^{+\infty} \frac{dq_0'}{2\pi}
\frac{\rho_{\mn}(q_0',\vec{q})}{q_0'-q_0-i\eta\epsilon(q_0)}
\label{diag_M}
\eea
where $\rho_{\mn} (q)$ is the spectral function
\bea
\rho_{\mn}(q) = \int d^4x e^{iq\cdot x}\la [J_\mu(x),J_\nu(0)]\ra \equiv
M^+(q) - M^-(q)
\label{spec_func}
\eea
Eq. (\ref{diag_M}) gives us 
\bea
\rho_{\mn}(q) = 2~\rm{Im}\ov{M}_{\mn}(q).
\label{rhotom}
\eea
From the cyclicity of the thermal trace, we get the Kubo-Martin-Schwinger
relation
\bea
M^+(q) = e^{\beta q_0}M^-(q)
\label{kms}
\eea 
From Eqs. (II.9-11) we get
\bea
M^+(q)=\frac{2e^{\beta q_0}}{e^{\beta q_0}-1}\rm{Im}\ov{M}_{\mn}(q)
\eea
giving the dilepton rate (\ref{dileptonrate}) as 
\bea
\frac{d^4N}{d^4q} = \frac{\alpha^2}{3\pi^3q^2}\frac{W}{e^{\beta q_0}-1},
~~W=-g^{\mn} \textrm{Im} \ov{M}_{\mn}.
\eea
Using the matrix which diagonalizes the $2\times 2$ correlation matrix, we can relate 
the imaginary part of any one component, say the $11$, of the correlation function to 
that of its diagonal element,
\bea
\textrm{Im} \ov{M}_{\mn} = \epsilon(q_0) \tanh (\beta q_0)\textrm{Im}(M_{\mn})_{11}.
\label{parl_matrix_comp}
\eea

So far we utilize general properties of two-point functions to relate the problem to 
$(M_{\mn})_{11}$. Taking $\tau$ and $\tau'$ on the real axis, the contour form 
(\ref{contourform}) gives it as 
\bea
M_{\mn}(x,x')_{11} = i\la TJ_\mu(\vec{x},t)J_\nu(\vec{x}',t')\ra,
\eea
where $T$ as usual time orders the operators. It is this quantity which we
have to calculate. To leading order in strong interactions, it involves only 
the thermal quark propagator. The magnetic field enters the problem through 
this propagator, which we find in the next section.

\section{Dirac propagator in magnetic field}
\label{Dirac_schwinger}

In deriving the quark propagator we assume both $u$ and $d$ quarks to have the same 
absolute electric charge as that of the lepton. (The necessary correction will be 
included in our formulae at the end of Section \ref{current_corr}). The Dirac Lagrangian 
in an external electromagnetic field 
\bea
\mathcal{L} = \bar{\psi}\left[i\gm^\mu(\partial_\mu+ieA_\mu)-m\right]\psi
\eea
gives the equation of motion
\bea
\left[i\gm^\mu(\partial_\mu+ieA_\mu)-m\right]\psi = 0.
\eea
Then the propagator 
\bea
S(x,x') = i\la 0 | T\psi(x)\bar{\psi}(x')| 0\ra
\eea
satisfies
\bea
\left[i\gm^\mu(\partial_\mu+ieA_\mu)-m\right]S(x,x') =-\delta^4(x-x').
\label{greens_function}
\eea
Here $ |0\ra$ is the vacuum state of the Dirac field (in presence of $A_\mu$). 
Defining states labeled by space-time coordinate (suppressing spinor
indices), we regard $S(x,x')$ as the matrix element of an operator $S$
\bea
S(x,x') = \la x | S |x'\ra.
\eea
Then Eq.(\ref{greens_function}) can be written as
\bea
(\gm^\mu\pi_\mu-m)S=-1,~~~  \pi_\mu=p_\mu-eA_\mu,~~~p_\mu=i\partial_\mu
\eea
which has the formal solution \footnote{Another equivalent form follows by
writing $(\slashed{\pi}+m)$ on the right in Eq.(\ref{formalsoln})
\cite{Schwinger}, but we shall not use it.}
\bea
S=\frac{1}{-\slashed{\pi}+m}=(\slashed{\pi}+m)\frac{1}{-\slashed{\pi}^2+m^2}.
\label{formalsoln}
\eea
Schwinger relates these quantities to the dynamical properties 
of a 'particle' with coordinate $x^\mu$ and canonical and
kinematical momenta $p^\mu$ and $\pi^\mu$ respectively. Using their
commutation relations, we get $\slashed{\pi}^2=\pi^2-\frac{e}{2}\sg F$ .
Defining $H=-\pi^2+m^2+\frac{e}{2}\sg F$ we can write Eq.(\ref{formalsoln}) as 
\bea
S=(\slashed{\pi}+m)i\int\limits_0^\infty ds U(s),~~~~ U=e^{-iHs}.
\eea
As the notation suggests, $U(s)$ may be regarded as the evolution operator of  
the particle with Hamiltonian $H$ in time $s$.

We now go to Heisenberg representation, where the operators $x_\mu$ and $\Pi_\mu$
as well as the base ket become time dependent,
\bea
x_\mu(s)=U^\dagger(s) x_\mu U(s),~~ \Pi_\mu(s)=U^\dagger(s) \Pi_\mu U(s),
~~~~|x';s\ra = U^\dagger(s)|x';0\ra
\label{heisenrep}
\eea
Then the construction of the propagator reduces to the evaluation of 
\bea
\la x''|U(s)| x'\ra = \la x'';s| x';0\ra,
\eea
which is the transformation function for a state, in which the operator $x_\mu(s=0)$ 
has the value of $x'_\mu$, to a state, in which $x_\mu(s)$ has the value $x''_\mu$. 

The equation of motion for $x_\mu(s)$ and $\pi_\mu(s)$ following from 
Eq.(\ref{heisenrep}) can be solved to get 
\bea
\pi(s)=-\frac{1}{2}eFe^{-eFs}\sinh^{-1}(eFs)\left(x(s)-x(0)\right)
\label{pis}
\eea
which may also be put in the reverse order on using the antisymmetry of $F_{\mn}$. 
The matrix element $\la  x^{\prime\prime};s| \pi^2(s)|  x^{\prime};0\ra$ can 
now be obtained by using the commutator $\left[x_\mu(s),x_\nu(0)\right]$ to reorder 
the operators $x_\mu(0)$ and $x_\nu(s)$. We then get
\bea
\la  x'';s| H(x(s),\pi(s))|  x';0\ra = f(x'';x';s)\la x'';s| x';0\ra
\eea
where
\bea
f\!\!=\!\!(x''-x')K(x''-x')\!\!-\!\!\frac{i}{2}\textrm{Tr}[eF\coth(eFs)]\!\!-\!\!m^2\!\!
-\frac{e}{2}\sg F,~~K\!\!=\!\!\frac{(eF)^2}{4}\sinh^{-2}(eFs).
\label{fform}
\eea

We are now in a position to find the transformation function, which from 
Eq. (III.10) is found to satisfy
\bea
i\frac{d}{ds}\la x'';s| x';0\ra = \la x'';s| H |x';s\ra
\eea
It can be solved as 
\bea
\la x'';s| x';0\ra &=& \phi(x'',x').\frac{i}{(4\pi)^2s^2}e^{-L(s)} \times \nn\\
&& \exp\left(-\frac{i}{4}(x''-x')eF\coth(eFs)(x''-x')\right)
\exp\left(-i(m^2+\frac{1}{2}e\sg F)\right)
\label{tfunc}
\eea
where
\bea
L(s)=\frac{1}{2}\textrm{Tr}\ln\left[(eFs)^{-1}\sinh(eFs)\right].
\eea
Here $\phi(x'',x')$ is a phase factor involving an integral over 
the potential $A_\mu$ on a straight line connecting $x'$ and $x''$. 
It will cancel out in our calculation. The spinor propagator is now given by
\bea
S(x'',x')&=& i\int\limits_0^\infty ds \la x''| (\slashed{\pi}+m)U(s)| x'\ra\nn\\
&=& i\int\limits_0^\infty ds \left[\gm^\mu\la x'';s| \pi_\mu(s)| x';0\ra
+m\la x'';s| x';0\ra\right]
\eea
with $\pi_\mu(s)$ and $\la x'';s| x';0\ra$ given by 
Eqs. (\ref{pis}) and (\ref{tfunc}).

We now specialize the external electromagnetic field to magnetic field $B$ in the 
$z$ direction, $F^{12}=-F^{21}=B$. It is convenient to diagonalize the antisymmetric 
$2\times 2$ matrix $F^{ij}$ with eigenvalues $\pm iB$. Going over to spatial metric 
we get \footnote{For any two vectors $a^\mu$ and $b^\mu$, we write
$(ab)_\pl=a^0b^0-a^3b^3$ and $(ab)_\pp=a^1b^1+a^2b^2$. Note that the 
longitudinal and transverse directions are defined with respect to the direction 
of magnetic field, not the collision axis of ions.}
\bea
S(x)=\frac{i}{(4\pi)^2}\int\frac{ds}{s}&&\frac{eB}{\sin(eBs)}\exp\left[
\frac{i}{4}x_\pp^2 eB\cot(eBs)-\frac{i}{4s^2}x_\pl^2-i(m^2+\frac{1}{2}e\sg F)s\right]\times\nn\\
&& \left[\left(\frac{1}{2s}(x\cdot \gamma)_\shortparallel + m\right)\left(\cos \phi - \gamma^1\gamma^2\sin \phi\right)-\frac{eB}{2\sin\phi}(x\cdot \gamma)_\perp
\right]
\eea
which can be Fourier transformed to 
\bea
S(p)&=&i\int\limits_0^\infty ds~e^{is(p^2-m^2+i\eps)}~e^{-isp_\pp^2\left(\frac{\tan(eBs)}{eBs}-1\right)}\times\nn\\
&& \left[(\slashed{p}_\pl +m)\left(1-\gm^1\gm^2\tan(eBs)\right)-\slashed{p}_\pp\left(1+\tan^2(eBs)\right)\right].
\eea
Expanding the exponential and tangent functions, we immediately get $S(p)$ as a 
series in powers of $eB$. To order $(eB)^2$ it is
\bea
S(p)=\frac{-(\slashed{p}+m)}{p^2-m^2+i\eta}+eB\frac{i(\slashed{p}_\pl+m)\gm^1\gm^2}{(p^2-m^2)^2}
-(eB)^2\left[\frac{2\slashed{p}_\pp}{(p^2-m^2)^3}-\frac{2p_\pp^2(\slashed{p}+m)}{(p^2-m^2)^4}\right].
\label{wfprop}
\eea

To put the propagator (\ref{wfprop}) in the form of a spectral representation, we 
introduce a variable mass $m_1$ to replace $1/(p^2-m^2)$ by $1/(p^2-m_1^2)$, 
keeping the physical mass $m$ unaltered at other places. The higher powers of the 
scalar propagator can then be expressed as derivatives of the propagator 
with respect to $m_1^2$. We thus get
\bea
S(p)=-F(p,m,m_1)\frac{1}{p^2-m_1^2}\Bigg|_{m_1=m},
\label{wfp_spec}
\eea
where 
\bea
F=(\slashed{p}+m)+a~i~\left(\slashed{p}_\pl + m\right)\gm^1\gm^2 + b\slashed{p}_\pp +cp_\pp^2 (\slashed{p}+m)
\eea
with coefficients $a, b$ and $c$ carrying the derivative operators,
\bea
a=-eB\frac{\partial}{\partial m_1^2};~~b=(eB)^2\frac{\partial^2}{\partial (m_1^2)^2};
~~c=-\frac{1}{3}(eB)^2\frac{\partial^3}{\partial (m_1^2)^3}.
\eea
From Eq.(\ref{wfp_spec}) the spectral function for $S(p)$ will be recognized as 
\footnote{We have three different spectral functions in this problem. The $\rho_{\mn}$ 
introduced in Section II is the spectral function of current correlation function, while
$\rho$ and $\sg$ are spectral functions for the scalar and Dirac propagators.}
\bea
\sg(p)=F(p,m,m_1)\rho(p,m_1)\big|_{m_1=m}
\eea
with $\rho$ being the spectral function for the scalar propagator of mass $m_1$ 
\bea
\rho(p,m_1)=2\pi\epsilon(p_0)\delta(p^2-m_1^2).
\eea
The desired spectral representation for the spinor propagator in vacuum 
(in presence of magnetic field) can be written in the form
\bea
S(p)=\int_{-\infty}^{+\infty}\frac{dp_0'}{2\pi}~\frac{\sg(p_0', \vec{p})}{p_0'-p_0-i\eta\eps(p_0)},
\eea
as can be readily verified by doing the $p_0'$ integral.

\section{Thermal current correlation function}
\label{current_corr}

\begin{center}
\begin{figure}[tbh]
\begin{center}
\includegraphics[scale=0.5]{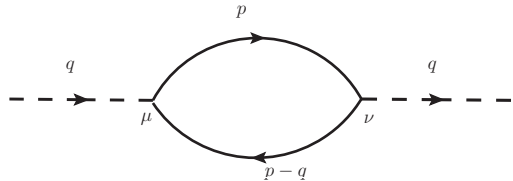} 
\caption{Current correlation function to one-loop. The dashed and solid lines 
represent currents and quarks.}
\label{cc}
\end{center}
\end{figure}
\end{center}

Like the thermal correlation function of currents, the thermal quark
propagator can also be analyzed in the same way. In particular, its
$2\times 2$ matrix form can be diagonalized, again with essentially a
single diagonal element, which turns out to be the vacuum propagator (in
magnetic field) derived above. But in our calculation below, we need the
$11-$ element of the original matrix, which is conveniently written as
\cite{Mallik},
\bea
S_{11}(p)=\int\limits_{-\infty}^{+\infty}
\frac{dp_0'}{2\pi}\sg(p_0', \vec{p})
\left\{\frac{1-\wt{f}(p_0')}{p_0'-p_0-i\eta}+\frac{\wt{f}(p_0')}{p_0'-p_0+i\eta}\right\},
~~\wt{f}(p_0) = \frac{1}{e^{\beta p_0}+1}
\eea
where the spectral function $\sg$ is given by (III.24). 

The graph of Fig. \ref{cc} gives two terms involving $u$ and $d$ quark propagators.
Assuming these to be equal (which is true only for $eB=0$)\footnote{For $eB 
\neq 0$, we include the necessary correction at the end of this section.}, we 
combine them to give
\bea
\left(M_{\mn}(q)\right)_{11} = \frac{5i}{3}\int
\frac{d^4p}{(2\pi)^4}\textrm{Tr}\left[S_{11}(p)\gm_\nu S_{11}(p-q) \gm_\mu\right]
\eea
where the prefactor includes a factor of 3 for color of the quarks.
Inserting the propagator (IV.1) in it, we want to work out the $p_0$
integral. For this purpose we write it as 
\bea
M_{\mn}(q)_{11}=\int\frac{d^3p}{(2\pi)^3}\int \frac{dp_0'}{2\pi}\rho(p_0',\vec{p})\int
\frac{dp_0''}{2\pi}\rho(p_0'',\vec{p}-\vec{q})K_{\mn}(q)
\label{p0integral}
\eea
where
\bea
K_{\mn}(q) &=&
i\int_{-\infty}^{+\infty}\frac{dp_0}{2\pi}N_{\mn}(q)\left(\frac{1-\wt{f}'}{p_0'-p_0-i\eta}
+\frac{\wt{f}'}{p_0'-p_0+i\eta}\right)\times\nn\\
&&
\left(\frac{1-\wt{f}''}{p_0''-(p_0-q_0)-i\eta}+\frac{\wt{f}''}{p_0''-(p_0-q_0)+i\eta}\right)
\eea
with $\wt{f}'=\wt{f}(p_0'), \wt{f}''=\wt{f}(p_0'')$ and
\bea
N_{\mn}(q)=\frac{5}{3} \textrm{tr}\left\{\stackrel{\lrw}{F}(p,m,m_1) \gm_\nu
\stackrel{\lrw}{F}(p-q,m,m_2)\gm_\mu\right\}.
\eea 
Here the masses $m_1$ and $m_2$ are variables on which the mass derivatives act in the 
two propagators. The left arrow on $F$ indicates the derivatives in it to be put 
farthest to the left (outside the integrals). As we are interested in the imaginary 
part of $K_{\mn}$, we can put $p_0=p_0',~ p_0-q_0=p_0''$ and bring $N_{\mn}$ outside 
the $p_0$ integral. Then it is simple to evaluate $K_{\mn}$, from which we get its 
imaginary part. Extracting a factor $coth(\beta q_0)$, it becomes linear in 
$\wt{f}'$ and $\wt{f}''$, 
\bea
\textrm{Im} K_{\mn}(q) = N_{\mn}(p_0',p_0'')\pi (\wt{f}''-\wt{f}')\coth(\beta q_0)\de
\left(p_0''-p_0'+q_0\right)
\eea
(The hyperbolic function will cancel out in (II.14)). Next, the $p_0'$ and 
$p_0''$ integrals in Eq.(\ref{p0integral}) can be removed, using the delta 
functions present in the spectral functions, namely $\de (p_0'\pm \om_1)$
and $\de (p_0''\pm \om_2)$ with $\om_1=\sqrt{|\vec{p}|^2+m_1^2}$ and
$\om_2=\sqrt{(\vec{p}-\vec{q})^2+m_2^2}$. We need only the imaginary part in 
the physical region, $q_0 > (\omega_1 +\omega_2)$. From (II.14), (IV.3) and
(IV.6), we then get
\bea
W=\pi\int\frac{d^3p}{(2\pi)^3}\frac{N_\mu^\mu(\om_1,-\om_2)}{4\om_1\om_2}
\left\{1-\wt{n}(\omega_1)-\wt{n}(\omega_2)\right\} \delta(q_0-\omega_1-\omega_2)
\eea
where we convert $\wt{f}$'s to distribution functions, $\wt{n}(\om)=1/(e^{\beta\om}+1)$.

Working out the trace over $\gm$ matrices in $N_\mu^\mu$ we get
\bea
N_\mu^\mu \!\!=\!\! -\frac{40}{3}\left[(1-a_1a_2)~p\!\cdot\!(p-q)\!-\!(b_1+b_2+a_1a_2)
[p\!\cdot\!(p-q)]_\pp \!\!+
\! p\!\cdot\!(p-q)\left\{c_1p_\pp^2+c_2(p-q)_\pp^2\right\}\right].\nn\\
\label{nmunu}
\eea
Let us now consider collision events in which the transverse components of momenta are 
small compared to the longitudinal ones, when we can omit the last two terms and 
calculate the dilepton rate analytically. Neglecting quark mass, we thus get
\bea
W=\frac{20\pi}{3} q^2(1-a_1a_2)J
\label{wtoj}
\eea
where 
\bea
J=\int\frac{d^3p}{(2\pi)^34\om_1\om_2}\left\{1-\wt{n}(\om_1)-\wt{n}(\om_2)\right\}\delta(q_0-\om_1-\om_2).
\label{j}
\eea
After working out this integral analytically, we shall apply the mass derivatives 
contained in $a_1$ and $a_2$.

If $\th$ is the angle between $\vec{q}$ and $\vec{p}$, we can carry out the $\th$ integral 
by the delta function in Eq.(\ref{j}). However a constraint remains to ensure that 
$\cos\th$ remains in the physical region, as we integrate over the angle. We get
\bea
J=\frac{1}{16\pi^2 |\vec{q}|}\int d\om_1 \Theta(1-| \cos\th|)
\left\{1-\wt{n}(\om_1)-\wt{n}(\om_2)\right\}.
\eea
The $\Theta$-function constraint gives a quadratic expression in $\omega_1$,
\bea
(\omega_1-\omega_+)(\omega_1-\omega_-)\le 0
\eea
where
\bea
\omega_{\pm}=\frac{q_0 R \pm |\vec{q}|\sqrt{R^2-4q^2m_1^2}}{2q^2},~~ R=q^2+m_1^2-m_2^2,
\eea
With the corresponding limits on $\omega_1$, we get~\cite{born}
\bea
J &=& \frac{1}{16\pi^2 |\vec{q}|}\int\limits_{\om_-}^{\om_+} d\om_1 
\left(1-\frac{1}{e^{\beta\om_1}+1}-\frac{1}{e^{\beta(q_0-\om_1)}+1}\right)\nn\\
&=& \frac{1}{16\pi^2 |\vec{q}|\beta} \left[\ln\left(\frac{\cosh(\beta\om_+/2)}
{\cosh(\beta\om_-/2)}\right)-\ln\left(\frac{\cosh(\beta(q_0-\om_+)/2)}
{\cosh(\beta(q_0-\om_-)/2)}\right)\right].
\eea

We now recall that the $u$ and $d$ quark charges were included correctly only in the currents but not in the propagators. The resulting correction will effect only $e^2$, contained in $a_1, a_2$ in the expression for $W$. We can readily find that we need to multiply $e^2$ by 17/45 to restore the actual charges of the quarks in their propagators.
Carrying out the mass derivatives in Eq.(\ref{wtoj}) and going to the limit of zero 
quark masses, we finally get 
\bea
W=\frac{5q^2}{12\pi |\vec{q}|\beta}\left[2\ln\left(\frac{\cosh \alpha_+}
{\cosh \alpha_-}\right)-\frac{17}{45}(eB)^2\mathcal{M}\right],
\eea
where $\mathcal{M}$ gives the effect of magnetic field to the leading order result,
\bea
\mathcal{M}=\frac{\beta^2}{8q^2}\left(\sech^2\alpha_+-\sech^2\alpha_-\right)+\frac{\beta|\vec{q}|}{q^4}\left(\tanh \alpha_+ +\tanh \alpha_-\right).
\label{mag_effect_final}
\eea
Here we use the abbreviation $\alpha_\pm=\beta(q_0\pm |\vec{q}|)/4$. Note that 
$W$ is finite as $|\vec{q}| \rightarrow 0$.

\section{numerical results and discussion}
\label{discussion}

Some earlier works estimate the magnetic contribution to dilepton rate
in QGP phase in heavy ion collisions. Ref. \cite{Tuchin1} uses 
Weizs$\ddot{a}$cker-Williams equivalent photon approximation, in which the 
two vertices of Fig.1 become independent amplitudes involving the photon, 
whose probabilities are calculated in the magnetic field. In Ref. {\cite{Sadooghi} 
the quark propagator is calculated using the method of eigenfunction expansion
 \cite{Ritus}. Here the anisotropy induced by the (constant) direction of the 
magnetic field is investigated in detail. In Ref. \cite{Bandyo} the result for 
very high magnetic field is reported, taking the lowest Landau level into account.  

Here we propose a different method to include the effect of magnetic field on 
the dilepton production rate. Assuming thermal equilibrium in the $QGP$ phase, 
there results the correlation function of quark currents. This is evaluated 
with the quark propagator in magnetic field after expanding it up to $(eB)^2$. 
The calculation is carried out in the real time method of thermal field theory.

The plots of $\mathcal{M}$, the coefficient of $(eB)^2$ in $W$, as functions 
of invariant dilepton mass $m_{l\bar{l}}=\sqrt{q^2}$ and temperature $T$ are 
shown in Fig. \ref{mwq} for typical values of parameters. If the second order 
term in (IV.16) provides any indication of the behavior of the series, the
expansion parameters are $eB/q^2$ and $eB/T^2$. In Fig. \ref{wr} we plot $W/W_{B=0}$ 
as a function of $m_{l\bar{l}}$ for a few values of $eB$.

\begin{center}
\begin{figure}[tbh]
\begin{center}
\includegraphics[scale=0.8]{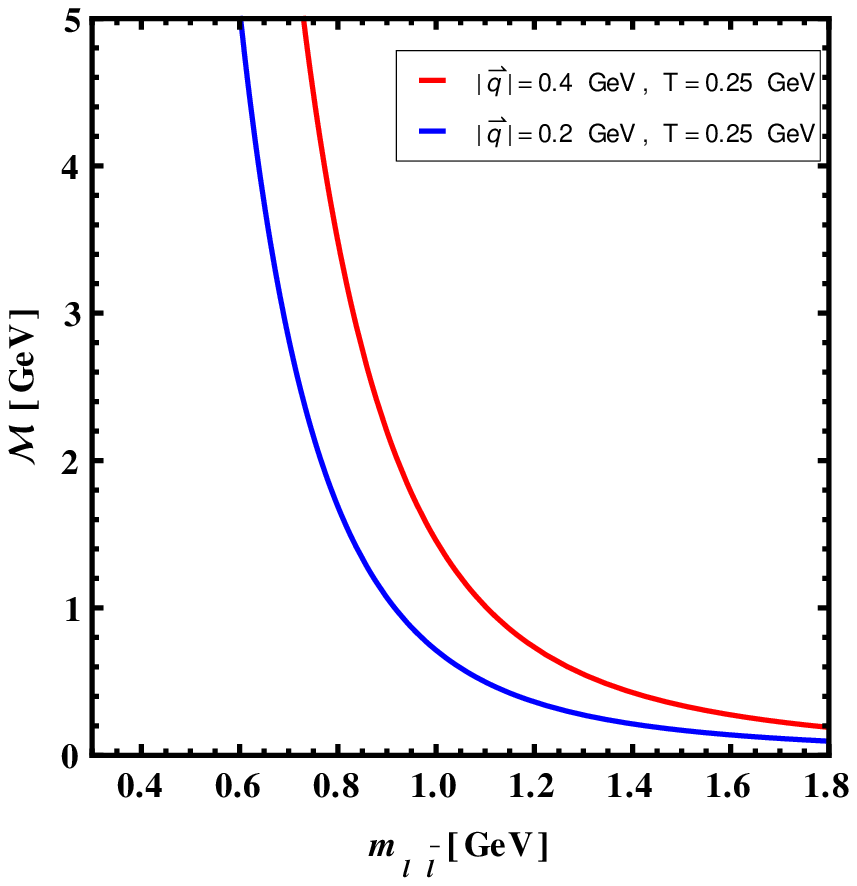}\includegraphics[scale=0.8]{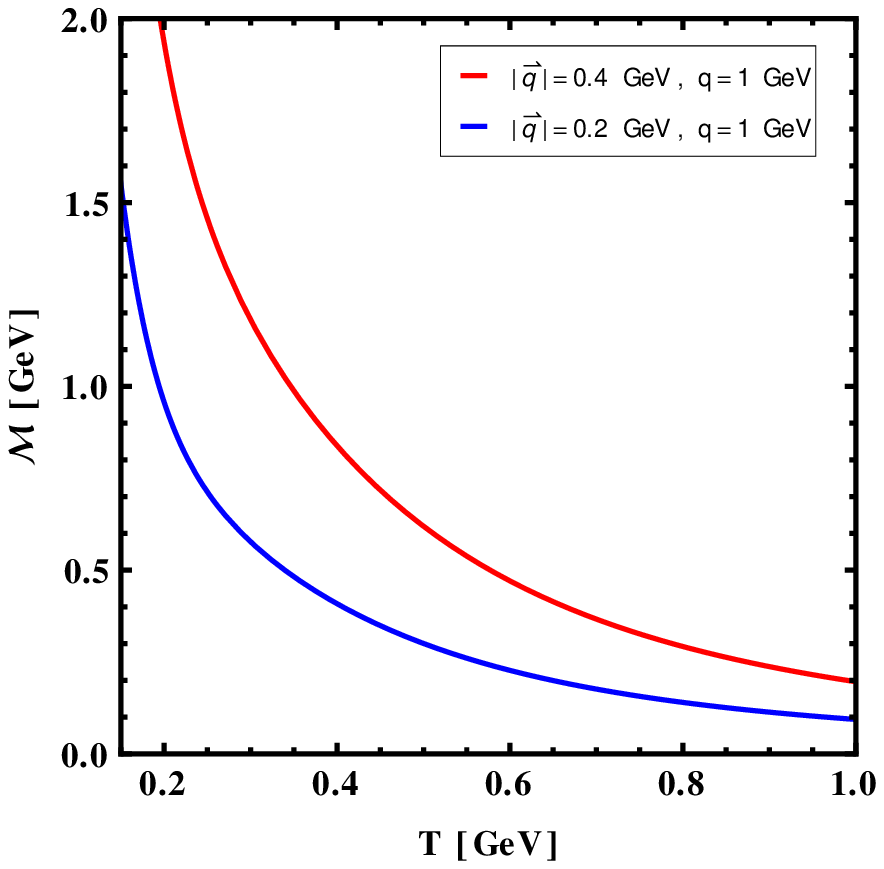}
\caption{Variation of $\mathcal{M}$ as a function of invariant mass
$m_{l\bar{l}}$ (left) and temperature $T$ (right).}
\label{mwq}
\end{center}
\end{figure}
\end{center}

\begin{center}
\begin{figure}[tbh]
\begin{center}
\includegraphics[scale=1.0]{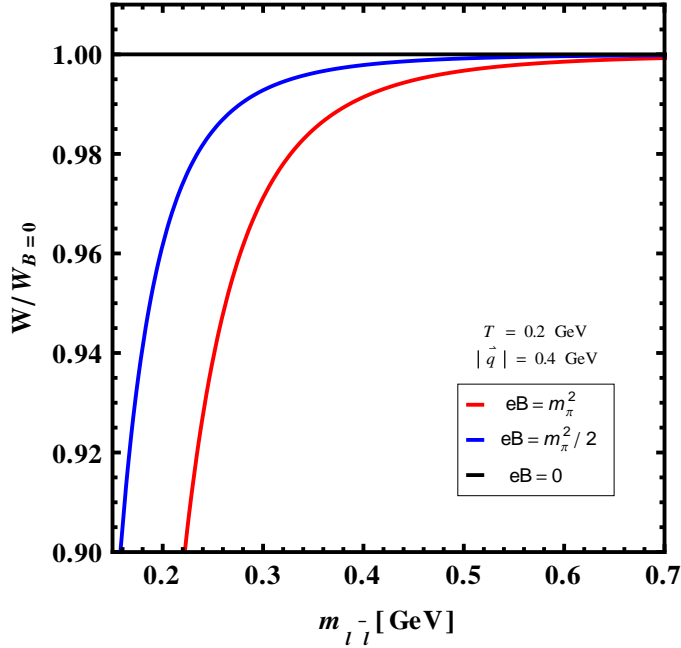}
\caption{Plot of the ratio of the dilepton rate with and without the presence of the magnetic
 field as a function of $m_{l\bar{l}}$.}
\label{wr}
\end{center}
\end{figure}
\end{center}

Fig. \ref{wr} shows that magnetic field changes the dilepton rate only at lower 
$q^2$, reflecting the behavior of the first and second term of
 $\mathcal{M}~(Eq.(\ref{mag_effect_final}))$ as $1/q^2$ and $1/q^4$ at low $q^2$.
 We also note that earlier theoretical calculations without magnetic field 
disagree with experiment at low $q^2$~\cite{ceres,helios,na38}. It would 
therefore be tempting to speculate if the effect of magnetic field can bring
the agreement, at least in part. However, to verify this speculation, we
have to improve our calculation in a number of ways. First, we should include
the terms in (IV.9) that are left out in our calculation. Then we need to
replace the constant magnetic field by one with its magnitude having
(adiabatic) time dependence, as realized in non-central collisions. One can also
include the first order $QCD$ correction to the current correlation function~\cite{Thoma}.

The time dependence of the magnetic field, mentioned above has to be included
 in the space-time evolution of dilepton production, which is needed to determine its spectrum.
 Without going into the details of this evolution, we may estimate roughly the effect of the
 time dependence as follows. The magnetic field realized in the core may be approximated
 as~\cite{Kharzeev,McLerran4,Tuchin1}
\bea
eB(t) = \frac{8\alpha}{\gamma} \frac{Z}{t^2+(2R/\gamma)^2}
\label{mag_field_td}
\eea
where $\alpha$ is the fine structure constant ($=1/137$), $Z$ and $R$ are the
 atomic number and radius of the colliding nuclei and $\gamma$ is the Lorentz
 contraction factor. This expression excludes large magnetic fields generated immediately 
 after collisions. So it may represent the magnetic field during the QGP phase.
Consider Au-Au collision at RHIC, for which $Z=79, R=6.5$ fm and $\gamma = 100$, 
giving $eB(t=0) = m_\pi^2/15$. However, considering Pb-Pb collision at LHC, where 
$Z=82, R=7.1$ fm and $\gamma = 2800$, we get $eB(t=0) = 1.4 m_\pi^2$. Concluding, we find that 
the effect of magnetic field in dilepton spectrum is confined to low invariant masses.

\end{document}